 \UseRawInputEncoding
\documentclass[aps,prl,twocolumn,epsfig,pdflatex]{revtex4-1}

\usepackage{amsmath,amssymb}
\usepackage{graphicx}

\usepackage{float}
\usepackage[usenames,dvipsnames]{xcolor}
\usepackage{amsthm,comment}
\usepackage{booktabs}
\usepackage[export]{adjustbox}
\usepackage[section]{placeins}
\usepackage[caption=false]{subfig}
\usepackage[percent]{overpic}
\bibpunct{[}{]}{,}{n}{}{}
\definecolor{red1}{HTML}{FF4136}
\definecolor{green1}{HTML}{00802b}

\usepackage[hidelinks]{hyperref} 

\hypersetup{colorlinks=true,citecolor=Red,linkcolor=Blue,urlcolor=Blue}
\usepackage[export]{adjustbox}

\graphicspath{{Figures/}}

\begin{document}

\title {Out of Equilibrium Majoranas in Interacting Kitaev Chains}
\author{Bradraj Pandey$^{1,2}$,
Narayan Mohanta$^{1,2}$, and Elbio Dagotto$^{1,2}$}
\affiliation{$^1$Department of Physics and Astronomy, University of Tennessee, Knoxville, Tennessee 37996, USA \\
$^2$Materials Science and Technology Division, Oak Ridge National Laboratory, Oak Ridge, Tennessee 37831, USA
 }

\begin{abstract}
We employ a time-dependent real-space local density-of-states 
method to study the movement and fusion of Majorana zero modes in the 1D interacting Kitaev model, based on the time evolution of many-body states.
 We analyze the dynamics and both fusion channels of  Majoranas using time-dependent potentials, either creating {\it Walls} or {\it Wells}.
For fast moving Majoranas, we unveil non-equilibrium signatures of the ``strong-zero mode'' operator  
(quasi parity degeneracy in  the full spectrum) and its breakdown in the presence of repulsive Coulomb interactions.
Focusing on forming a full electron after fusion, we also discuss upper and lower limits on 
the Majorana speed needed to reduce non-adiabatic effects and to avoid poisoning due to decoherence. 
\end{abstract}

\pacs{71.30,+h,71.10.Fd,71.27}

\maketitle

{\it Introduction.} Majorana zero modes (MZMs) generate considerable interest because of
potential applications in quantum information and computation~\cite{Kitaev,Freedman,Bonderson}.
MZMs obey non-abelian exchange statistics. Because they are
topologically protected from local perturbations and disorder, they are of value as possible qubits~\cite{Nayak,Mohanta,sarma}. 
Signatures of MZMs are expected to develop in tunneling conductance experiments
as zero bias peaks~\cite{Demler,Deng,Wang}. 
The simplest setup to realize Majoranas are quantum wires, 
where MZMs develop at the two edges~\cite{Alicea,Kitaev}. For ferromagnetic atomic chains 
with strong spin-orbit coupling placed over a superconductor, 
MZMs were indeed reported at the edges 
in spatially and spectral-resolved scanning tunneling experiments~\cite{Yazdani,jeon}.  
In nanowires, most theoretical work neglect repulsion among particles. 
However, Coulomb repulsion
plays an important role in 1D MZMs because 
it suppresses the pairing-induced bulk gap and can
destroy topological protection~\cite{Simon,Peter}. 
 
The movement of Majoranas and detection of fusion channels are important for  
quantum-information processing~\cite{Alicea,Aasen}. 
MZMs behave as Ising non-Abelian anyons~\cite{Nayak,Grosfeld} and obey the fusion 
rule~\cite{Aasen}: $\gamma \times \gamma= I + \psi $, 
meaning two MZMs can fuse into the vacuum $I$
 or into an electron $\psi$. The fusion process requires a slow adiabatic movement 
of Majoranas, achieved by applying properly adjusted time-dependent 
local gates to the topological superconducting wire~\cite{Alicea}. 
The rapid progress in quantum-wires with tunable local 
gates~\cite{Dartiailh,Aasen,Alicea}  provides a promising platform for 
creation, movement, and fusion of Majoranas~\cite{Zhou}.

Motivated by experimental progress in nanowires~\cite{Zhang}, here  
for the first time we employ a computationally-intensive time-dependent
real-space local density-of-states $LDOS(\omega,j,t)$ method to
observe the movement and fusion of Majoranas in the interacting Kitaev model. 
The uniqueness of our effort is that we can study Majorana movement at {\it any} speed by properly choosing the time dependence of gate voltages, namely we
can access the non-equilibrium situation away from adiabaticity, difficult to reach by theoretical tools.
$LDOS(\omega,j,t)$ of moving Majoranas and fusion outcome can be 
 measured in tunneling spectroscopy experiments based on gate-control nanowire devices~\cite{Zhang,Chetan2}.
Compared to previous studies based on single particle states, here
we use the exact-diagonalization method for the time evolution of the many-body 
states of {\it interacting} electrons in the 1D Kitaev model up to 16 sites.
We address the out-of-equilibrium properties  and fusion rules of MZMs, via sequential application of time-dependent chemical potential gates. 
For fast moving non-interacting MZMs, 
using the time-dependent LDOS,
we find the signature of a ``strong-zero mode'' operator in $LDOS(\omega,j,t)$~\cite{Kells,Fendley}. 
Remarkably, we found the total spectral weight at $\omega=0$ remains conserved (almost identical to the case of slow moving Majoranas).
However, with interaction $V$, depending on its strength and switching time $\tau$, 
we find loss in spectral weight at  $\omega=0$ and  breakdown  of the strong-zero mode properties. 
Furthermore, we provide the  time scale to observe the fusion rules of interacting Majoranas.
Although it is widely expected that for ``adiabatic" movement of MZMs their fusion will lead
to the formation of a fermion or vacuum states, only via calculations as presented  here, 
that allows for {\it any} speed for the MZMs, 
is that we can establish how ``slow" the movement must truly be in practice.

{ \it Model and Method.} 
We consider the time-dependent interacting 1D Kitaev model for spinless fermions 
 with open boundary conditions to anchor MZMs:  
\begin{eqnarray}
H(t) = -t_h \sum_{i=1}^{N-1} \left( c^{\dagger}_{i}c^{\phantom{\dagger}}_{i+1}+H.c.\right)  
+V \sum_{i=1}^{N-1} \left(n_i n_{i+1}\right)+ \nonumber 
\end{eqnarray}
\begin{eqnarray}
\Delta \sum_{i=1}^{N-1} \left(c^{\dagger}_i c^{\dagger}_{i+1} + H.c. \right) +\sum_{i}^{N} \mu_{i}(t)n_i, 
\end{eqnarray}
\noindent where $n_i= c_i^{\dagger}  c_i^{\phantom{\dagger}} $ and $c_i^{\dagger}$ ($c_i^{\phantom{\dagger}}$) 
is the fermionic creation (annihilation) operator, 
$t_h$ is the hopping amplitude, and $\Delta$ is the 
$p$-wave pairing strength. The time dependent Hamiltonian $H(t)$ (Eq.1) commutes with the parity operator $P=e^{i\pi\sum_jn_j}$~\cite{Turner,Miao}.
The time dependence is incorporated in the 
chemical potential $\mu_i(t)$ as:
\begin{figure}[h]
\centering
\rotatebox{0}{\includegraphics*[width=\linewidth]{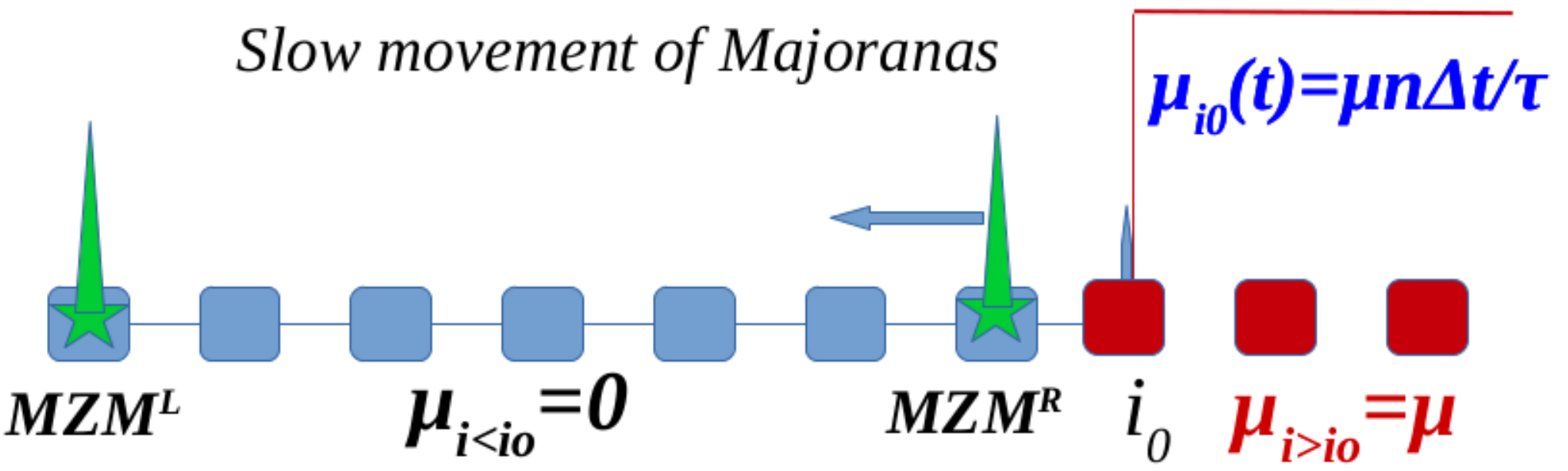}}
	\caption{Schematic representation of the transfer of 
	the right edge MZM$^{R}$  to site $i_0-1$ 
        in a 1D Kitaev chain. The onsite chemical potential at 
        any site can be tuned using time-dependent local gates with 
        quench rate $1/\tau$. Large (small) $\tau$ corresponds to slow (fast) motion of MZM$^R$.
	Blue squares denote the topological region 
        (with $\mu_{i<i_o}=0$) while red squares denotes the non-topological region 
        (with $\mu_{i>i_o}=\mu$) for this {\it Wall} case. For the {\it Well}, the red region has a negative $\mu$.     
	}
\label{Fig1}
\end{figure}
\begin{eqnarray}
\mu_{i}(t)=0  \  (i <i_0), \nonumber
\ ~~~~\mu_{i}(t)=\mu \ (i > i_0), 
\end{eqnarray}
\begin{equation}
 \mu_{i}(t)=\mu\frac{n\Delta t}{\tau} \ ( i=i_0), 
\end{equation}
\noindent where $1/\tau$ is the quenched rate, $\Delta t= 0.001$ is the small time step we used,
and $n$ is the integer number of those steps, such that 
the on-site chemical potential $\mu_{i}(t)$ at $i=i_0$ 
increases approximately linearly from 0 to $\mu$ in a time $\tau$ 
(defined as the switching time of the local gate at site $i=i_0$). 
The sequential application of onsite gates $\mu_i(t)$ on the right half of the 1D chain, creates a 
moving {\it Wall} for $\mu>0$ (or moving {\it Well} for $\mu<0$), separating topological from non-topological 
regions at site $i=i_0$.
Equating our number of sites with number of gates in a coarse-grained approach, 
 this process leads to the movement of  the right edge Majorana zero mode (MZM$^{R}$) 
from the edge $i=N$ to site $i_0-1$ in a finite time 
$t=N_R \tau$, with $N_R$ the number of sites where the chemical potential reached
its maximum value (here being $|\mu|=12$) at time $t$ (Fig.~\ref{Fig1}).      

To calculate the time-dependent local density-of-states (at zero temperature), 
we first time evolve the ground-state wave function $|\psi(0)\rangle$ up to time $t=N_R\tau$, 
	using the time-dependent Hamiltonian $H(t)$ as: $|\Psi(t)\rangle= \mathcal{T} exp{\left (-i\int_{0}^{t}H(s) ds\right)}|\psi(0)\rangle$, 
	where $\mathcal{T}$ is the time ordering operator~\cite{Sen}.  
	Then, we calculate the double-time Green function $G(t,t')$~\cite{Kennes},
        using the time-independent Hamiltonian $H_f = H(t=t_f)$ at time $t_f=N_R \tau$:
\begin{equation}
G^{elec}_j(t,t')= \langle \Psi(t)|c^{\dagger}_j e^{iH_ft'}c_j e^{-iH_ft'}|\Psi(t)\rangle.
\end{equation}
The time-dependent $LDOS_{elec}$($\omega,j,t$) for electrons is the 
	Fourier transform with respect to $t'$  of the local-Green function at site $j$: $LDOS_{elec}$($\omega,j,t$)= 
\begin{equation}
=\frac{1}{\pi}Im\int_0^{T} dt' e^{i \left(\omega+i \eta\right) t'} iG^{elec}_j(t,t'). 
\end{equation}
\noindent where we use $T=70$ for the integration, and broadening $\eta=0.1$. Similarly we obtained the $LDOS_{hole}$($\omega,j,t$) for holes, using the Fourier transform of the Green function $G^{hole}_j(t,t')=\langle \Psi(t)|c_j(t')c^{\dagger}_j|\Psi(t)\rangle $. The total local density-of-states at site $j$ is, thus, $LDOS(\omega,j,t) =LDOS_{hole}(\omega,j,t)+LDOS_{elec}(\omega,j,t)$.
 
\begin{figure}[h]
\centering
\rotatebox{0}{\includegraphics*[width=\linewidth]{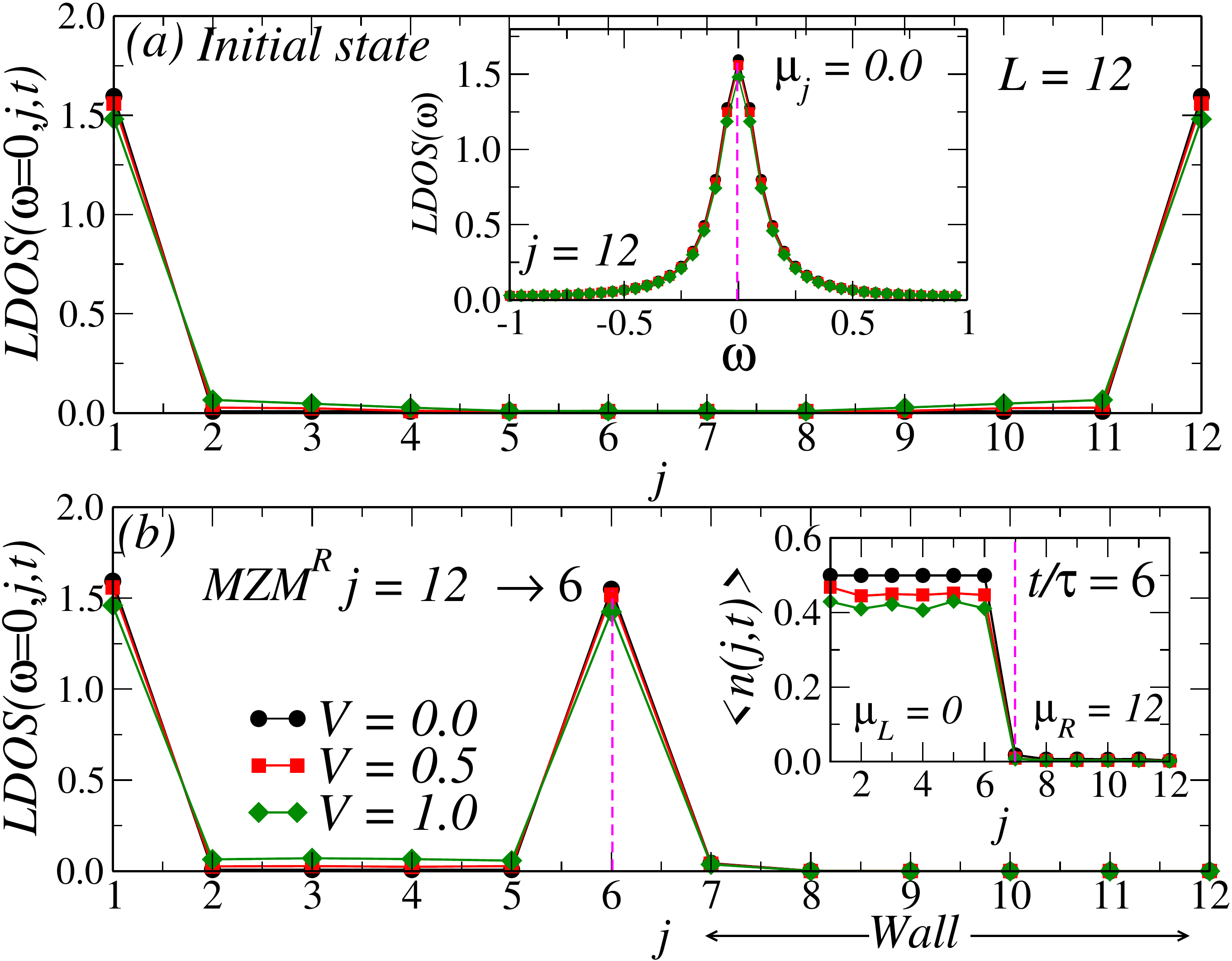}}
\rotatebox{0}{\includegraphics*[width=\linewidth]{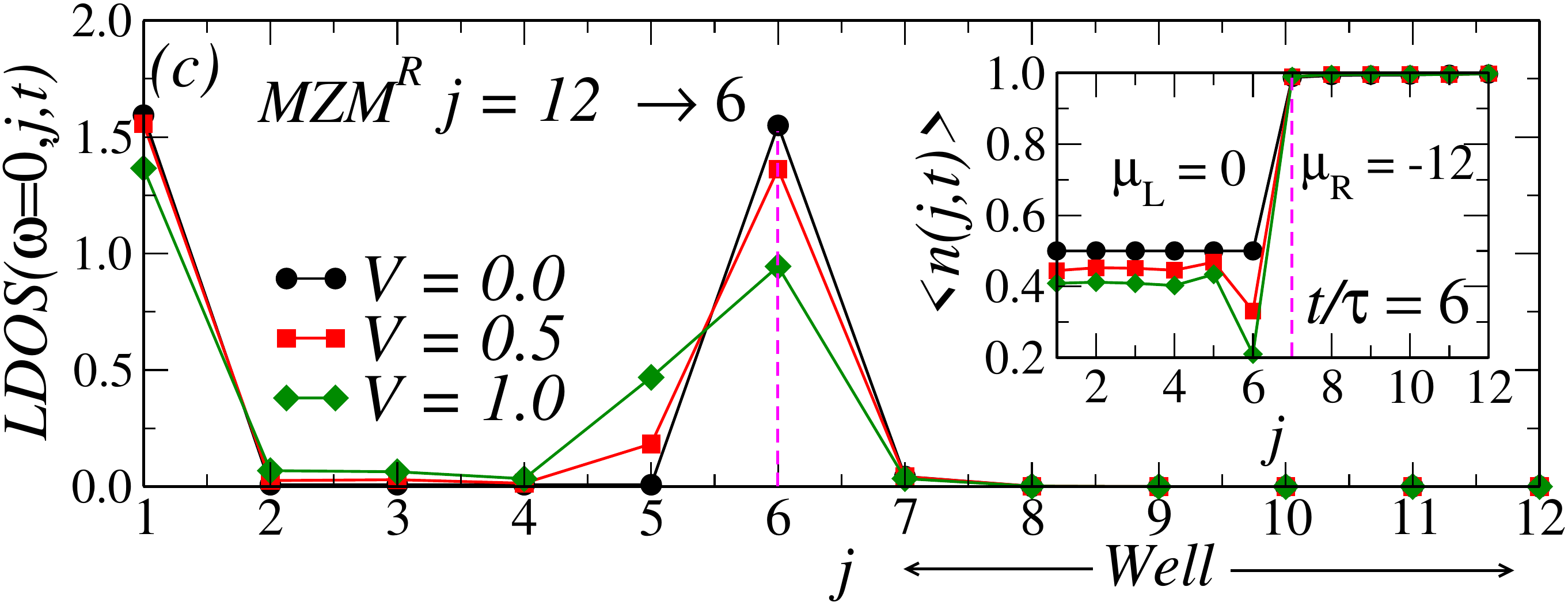}}
	\caption{
	Slow movement of Majoranas. (a) Local density-of-states  $LDOS(j,\omega,t)$ vs.
	site $j$, at time $t=0$ and for $\omega=0$.  
	The sharp peaks at sites $j=1$ and $j=12$ in
	$LDOS(j,\omega=0,t)$  represent Majorana edge modes for 
	different values of $V$ and $\mu_j=0$ (for $1<j<12$). 
        Inset: $LDOS(\omega)$ vs. $\omega$ at site $j=12$ using broadening $\eta=0.1$.
	 Site-dependent $LDOS(j,\omega,t)$ at $\omega=0$ and time
        $t/\tau=6$ for $V=0, 0.5$, and $1.0$ with $\tau=36$, $60$, and $72$, respectively, for:
	 (b) Positive $\mu$ ({\it Wall}). Inset shows the 
         site-dependent density $\langle n(j,t)\rangle$ at $t/\tau=6$ for $\mu_R=12$.
         (c) Negative $\mu$ ({\it Well}). Inset shows the
         site-dependent density $\langle n(j,t)\rangle$ at time $t/\tau=6$ 
	 for $\mu_R=-12$. $L=12$ sites and $t_h=\Delta=1.0$ were used.
	}
\label{Fig2}
\end{figure}
\begin{figure}[h]
\centering
\rotatebox{0}{\includegraphics*[width=\linewidth]{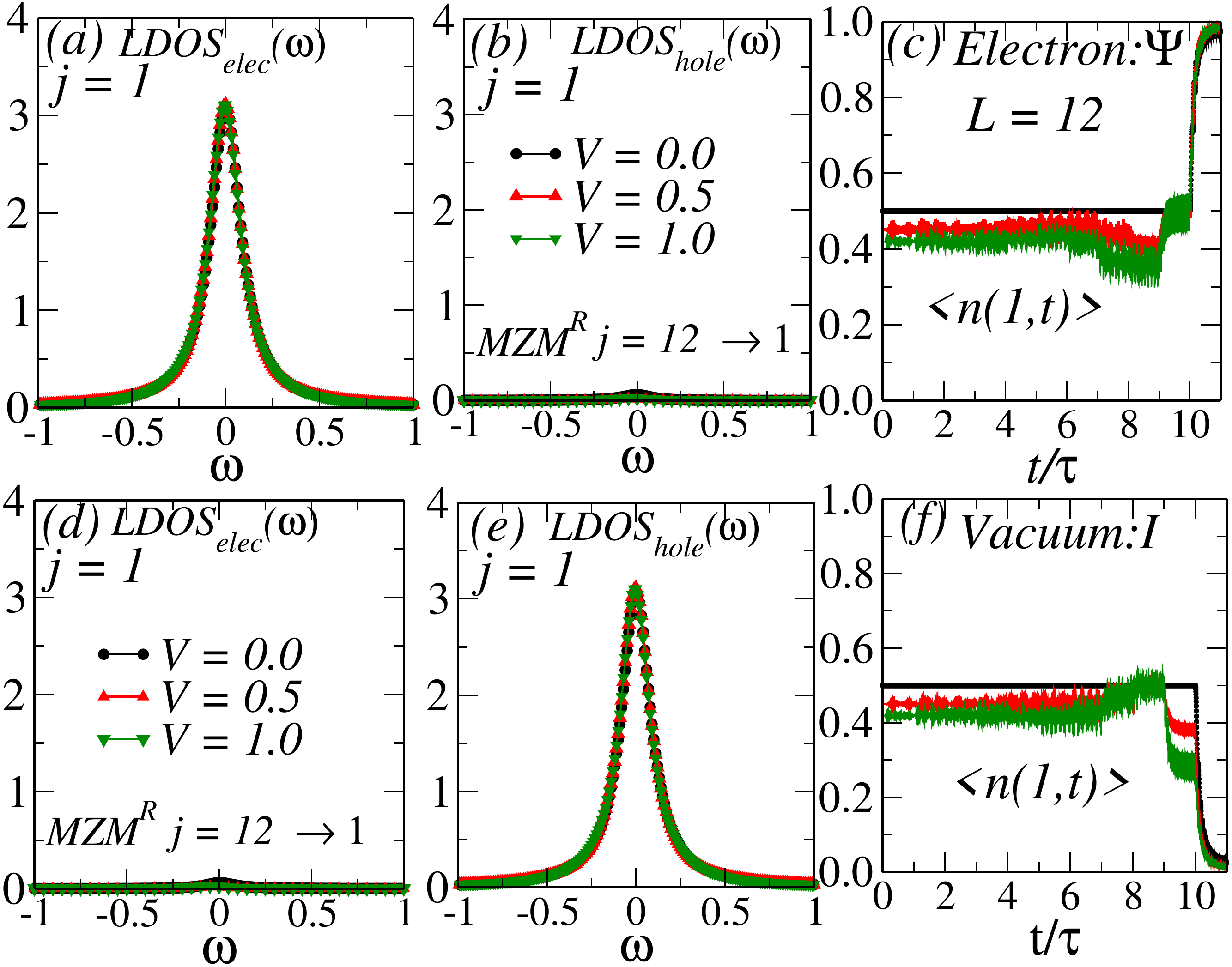}}
	\caption{Slow Majorana fusion  using $\mu(t)>0$ ({\it Wall}), $\Delta=1$, at $V=0.0$, $0.5$, 
	and $1.0$,  with $\tau=36$, $60$, and $72$, respectively.
        Upper panels show Majoranas fusion for the initial state with parity $P=-1$. 
	(a) Electron $LDOS_{elec}$($\omega$) after moving a MZM from $j=12$ to $j=1$.
	(b) Hole  $LDOS_{hole}$($\omega$) at $t/\tau=11$ and site $j=1$. 
	(c) Charge density $\langle n(j=1,t)\rangle$ vs. time $t$, varying $V$.
        Lower panels show Majoranas fusion for the initial state with parity $P= +1$.  
	(d) Electron $LDOS_{elec}$($\omega$) at $t/\tau=11$ and site $j=1$, 
	(e) Hole $LDOS_{hole}$($\omega$) after moving a MZM from $j=12$ to $j=1$.
	(f) Charge density $\langle n(j=1,t)\rangle$ vs. time $t$. 
	}
\label{Fig3}
\end{figure}
{ \it Slow movement of Majoranas.} For fusion or braiding of  MZMs, it is required to transfer the Majoranas slowly, close to the adiabatic limit~\cite{Karzig,Carrasquilla}. 
Figure~\ref{Fig2}(a) shows the real-space local density-of-states $LDOS(\omega=0,j,t=0)$
 vs. site $j$, with $\mu_i=0$ (for all sites), and at $t_h=\Delta=1$. 
For small or zero $V$, these peaks are sharply localized at the
 end sites ($i=1$ and $12$), whereas for robust $V$ the $\omega=0$ peaks are slightly delocalized over a few sites. 
In the inset, we show $LDOS(\omega)$ at time $t=0$ at the end site $j=12$ and several $V$'s. 
We  find a sharp peak at $\omega=0$, signaling a MZM mode at the end site. 
Integrating in $\omega$ the LDOS($\omega$,$j=12$) at $V=0.0$ gives spectral weight $~0.48$, 
close to the analytically expected value 0.5~\cite{foot}.

Next, with sequential application  of the time-dependent chemical potential $\mu_j(t)$, 
the right edge MZM$^{R}$ (at site $j=12$) is moved to the middle site ($j=6$)
 in a time $t=N_R \tau$ (i.e. $t/\tau=N_R=6$ because we travel 6 sites). We study cases $\tau=36, 60 $, and $72$, for
interaction strengths $V=0.0$, $0.5$, and $1.0$, respectively. In this case,
$|\psi(t)\rangle$ remains close to the degenerate ground-state space
(larger values of $V$ require a slower-rate of increase in the onsite $\mu_{i_0}(t)$).
As shown in Fig.~\ref{Fig2}(b), for $\mu=12$ (i.e. when creating a potential {\it Wall}),
the $LDOS(\omega=0,j,t)$ has peaks at sites $j=1$ and $j=6$  at
time $t/\tau=6$, indicating that a slow transfer of MZM$^R$ from $j=12$ to $j=6$ occurred.
The average density $\langle n(j,t) \rangle$ 
is close to zero for $j\ge 7$, while it is close to 0.5 
for $j\le 6$ (inset of Fig.~\ref{Fig2}(b)).
Interestingly, at $\mu_R=-12$  (when creating a potential {\it Well}), the effect of interaction increases. 
In the non-topological region ($j\ge$ 7), each site is occupied by one fermion, 
whereas in the topological region ($j\le$ 6) the mean occupancy is close to $0.5$. With nonzero $V$, to minimize the Coulomb interaction between 
fermions at the topological to non-topological boundary,
the fermions near the boundary become inhomogeneously distributed (Fig.~\ref{Fig2}(c), inset). 
This delocalizes MZM$^{R}$  
over more sites as $V$ increases (Fig.~\ref{Fig2}(c)).

{ \it Slow fusion of Majoranas.} For the fusion of Majoranas, 
we move the right edge MZM$^R$ slowly all the way to the left end (site $j=1$) using sequential 
operations of $\mu(t)$ in a time interval $t=11\tau$ (see caption Fig.~3). At $t=0$, for $V=0$, $t_h= \Delta=1$,
with $\mu_i=0$ (for all sites), the system has  degenerate many-body ground states ($|\psi_1\rangle$ and $|\psi_2\rangle$).
These degenerate ground states have different fermionic parity $P=\pm 1$.
At $t=0$, we start the time evolution with those initial states $|\psi_s\rangle$ ($s=1$ or $2$) up to
$t/\tau=11$, to confirm both fusion channels ($Electron:\Psi$ and $Vacuum:I$). 
For positive chemical potential, $\mu(t) >0$ ({\it Wall}) and the initial states $|\psi_s\rangle$  with parity $P=-1$, 
 the electron $LDOS_{elec}$($\omega$) at $j=1$
 shows a sharp peak close to $\omega=0$, for $V=0$, $0.5$, and $1$ (Fig.~\ref{Fig3}(a)).
 Meanwhile, the hole $LDOS_{hole}$($\omega$) at $j=1$, displays no peak (Fig.~\ref{Fig3}(b)). 
 The time-dependent density $\langle n(j=1,t) \rangle$ at site $j=1$ takes
 values one at $t/\tau=11$, giving a clear indication
 of the formation of a single electron (spinless)  
 at site $j=1$ after Majoranas fusion  (Fig.~\ref{Fig3}(c)).
On the other hand, for the initial state $|\psi_s\rangle$ 
with parity $P= +1$,
the hole $LDOS_{hole}$($\omega$) displays a sharp peak
 close to $\omega=0$, for $V=0$, $0.5$, and $1$ (Fig.~\ref{Fig3}(e)). The electron 
 $LDOS_{elec}$($\omega$) has no peak at $t/\tau=11$, see Fig.~3(d).
 The density $\langle n(j=1,t)\rangle$ at $j=1$ approaches zero (Fig.~\ref{Fig3}(f)), 
confirming a vacuum state at $t/\tau=11$.
Because the MZM spreads over more than one site
as $V$ grows, density fluctuations occur at site $j=1$ as
 compared to $V=0$ (Figs.~\ref{Fig3}(c,f)).

{ \it Fast movement of Majoranas.}
Changing $\mu(t)$ employing a faster rate (smaller $\tau$), leads to a fast movement of MZMs$^R$  generating non-adiabatic effects~\cite{Scheurer,Conlon}.
The faster change in $\mu(t)$ results in a finite overlap of the time-evolving wave-function $|\psi(t)\rangle$ with excited states
 of the instantaneous Hamiltonian $H(t)$.
Starting the time evolution with initial states 
$|\psi_s\rangle$, with parity $P=-1$ for $\mu(t)>0$ ({\it Wall}) and using all eigenvectors $\{|n\rangle \}$ of the instantaneous Hamiltonian $H(t)$ 
 the electron $LDOS$  at finite time $t=N_R \tau$ can be written as:  
$LDOS_{elec}(\omega,t)=$ 
\begin{eqnarray}
\frac{-1}{\pi} Im\left( \sum_{m,n} \frac{\langle \Psi(t)| c^+_j |n\rangle \langle n|c_j|m\rangle \langle m|\Psi(t)\rangle }{e_n-e_m+\omega+i \eta} \right),
\end{eqnarray}
where $\eta=0.1$, and the rest of the notation is standard (as reference for $L=12$ the number of states is 4,096).
\begin{figure}[h]
\centering
\rotatebox{0}{\includegraphics*[width=\linewidth]{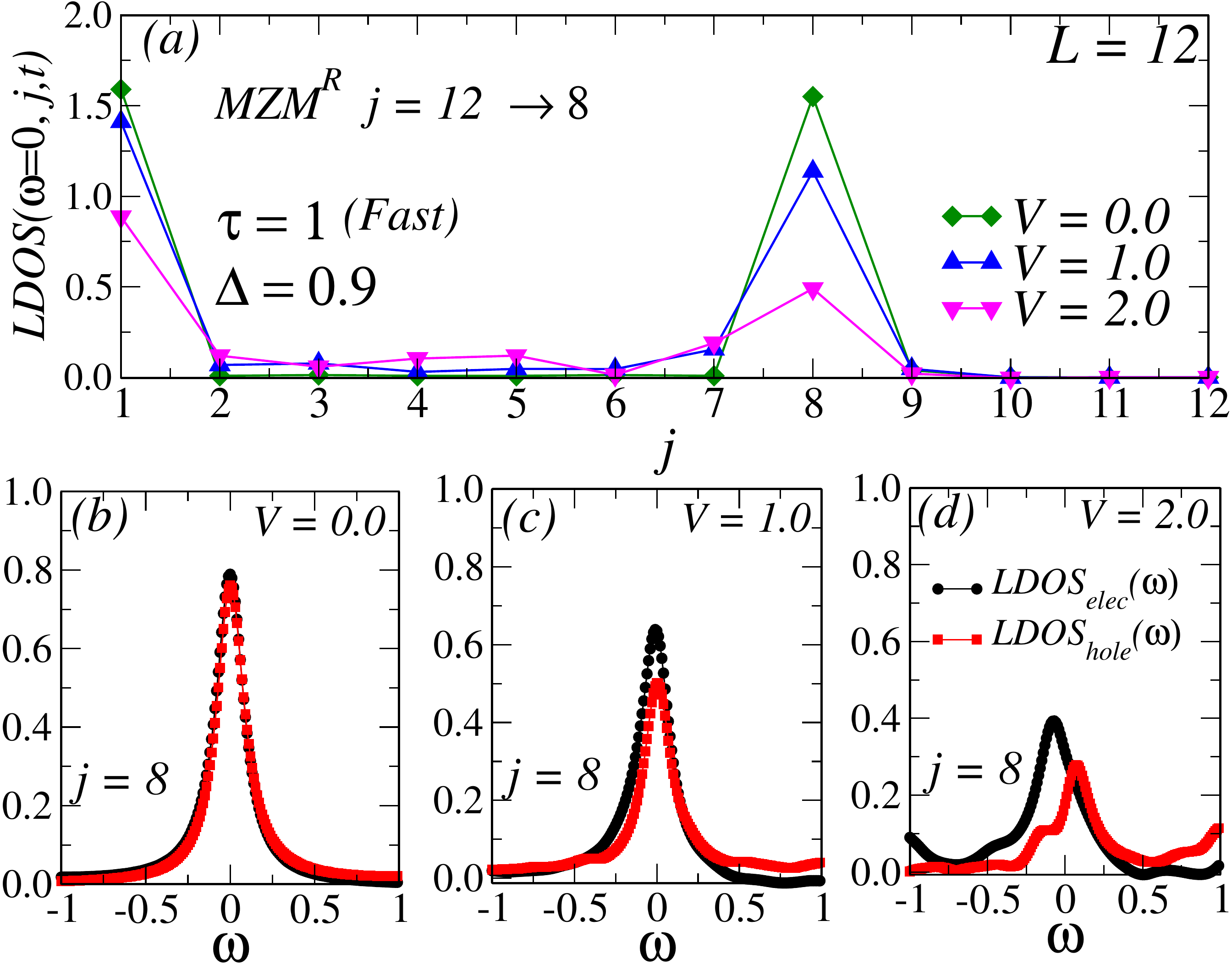}}
	\caption{Fast movement of Majoranas using a small quench rate $\tau=1$, $\mu(t)>0$ ({\it Wall}),
and for the initial state with parity $P=-1$.
(a) Site-dependent $LDOS(j,\omega,t)$ at $\omega = 0$, for $V$ = 0.0, 1.0, and 2.0,
after moving the right MZM from site 12 to 8. 
Electron $LDOS_{elec} (\omega)$ and Hole $LDOS_{elec}(\omega)$ at site $j=8$, 
for (b) $V=0.0$, (c) $V=1.0$, and (d) $V=2.0$.
	}
\label{Fig4}
\end{figure}
\begin{figure}[h]
\centering
\rotatebox{0}{\includegraphics*[width=\linewidth]{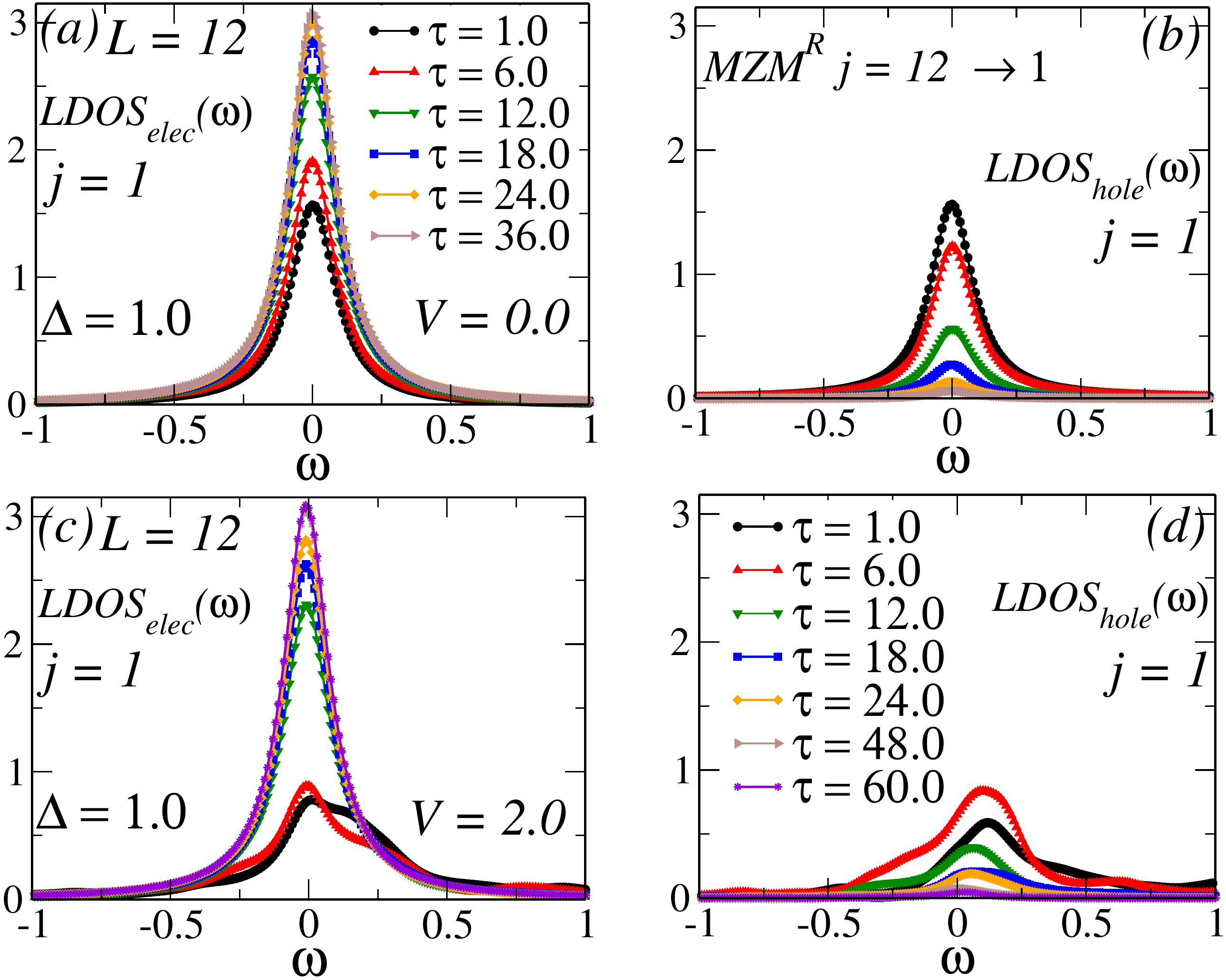}}
\rotatebox{0}{\includegraphics*[width=\linewidth]{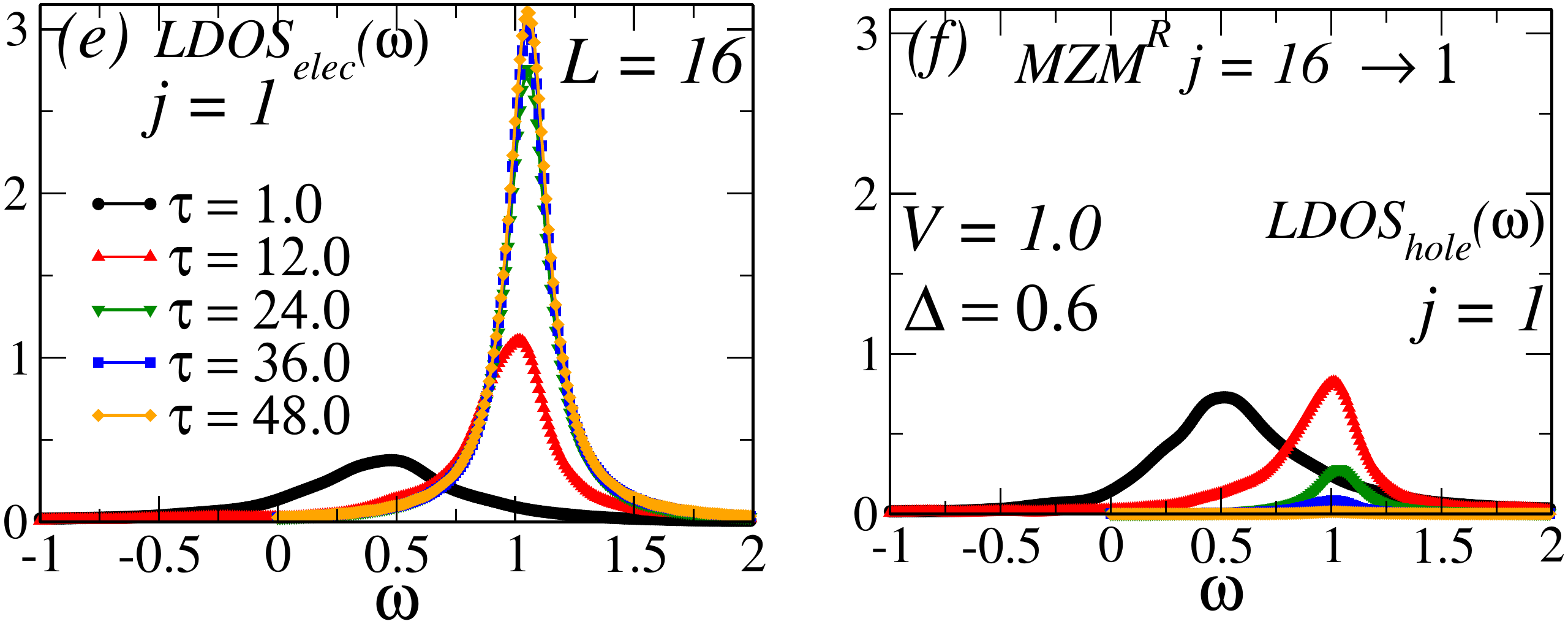}}
	\caption{
        Fast Majorana fusion using $\mu(t)>0$ ({\it Wall}) 
	at various $\tau$'s. The right MZM is moved from site $j=12$ to 1. 
        (a,b,c,d) Initial state with $P= -1$. 
	Panel (a) $LDOS_{elec}$($\omega$) at site $j=1$ and 
        (b) $LDOS_{hole}$($\omega$) at $j=1$, for $V=0$, $\Delta=1.0$, and $L=12$.
        Panel (c) $LDOS_{elec}$($\omega$) at site $j=1$ and 
        (d) $LDOS_{hole}$($\omega$) at $j=1$, for $V=2$, $\Delta=1.0$, and $L=12$.
        Fast Majorana fusion using $\mu(t)<0$ ({\it Well}) 
        at various $\tau$'s. The right MZM is moved from site $j=16$ to 1. 
        (e,f) Initial state with $P= +1$.
	Panel (e) $LDOS_{elec}$($\omega$) at site $j=1$ and 
        (f) $LDOS_{hole}$($\omega$) at $j=1$, for $V=1$, $\Delta=1.0$, and $L=16$.
	  } 
\label{Fig5}
\end{figure}

Figure~\ref{Fig4}(a) shows $LDOS(\omega,j,t)$ after moving the MZM from $j=12$ to $j=8$ for
different values of Coulomb interaction $V$ and with fast quench rate $\tau=1$. For $V=0$, and for this faster change in $\mu(t)$ 
the peak values of $LDOS(\omega,j,t)$ almost remain the same as compared to slower changes in $\mu(t)$ (Fig.~\ref{Fig2}(b)).
 On the other hand for the same faster moving MZMs but for a finite repulsion $V$,
 there is a significant reduction in the peaks magnitude of $LDOS(\omega,j,t)$ 
(at sites $j=1$ and $j=8$)  with increasing $V$. For larger values of $V$, 
the site-dependent $LDOS(\omega,j,t)$ indicate finite overlap between left and right MZMs (Fig.~\ref{Fig4}(a)). 
Figures~\ref{Fig4}(b,c,d) contains the electron $LDOS_{elec}(\omega)$ and hole  $LDOS_{hole}(\omega)$  vs. $\omega$ 
at site $j=8$, for different values of the Coulomb interaction $V$ at time $t/\tau=4$.
Using full-diagonalization of  $H(t)$, we find for $V=0$ and $\Delta=0.9$~\cite{foot2}
that all many-body eigenstates of $H(t)$ come in pairs with opposite parity $P= \pm 1$  
and the states of each pair are almost degenerate (the parity degeneracy of all many-body states is compatible with
a strong zero mode operator~\cite{Kells,Fendley}. Previous work showed that a zero mode operator commutes with the Hamiltonian up to exponentially small 
 finite-size corrections, leading to global parity degeneracy in the full spectrum~\cite{Fendley}). 
 The equal peak heights of $LDOS_{elec}(\omega,t)$ and $LDOS_{hole}(\omega,t)$ 
at $\omega=0$ in Fig.~\ref{Fig4}(b) (with spectral weight ~0.5), 
can be associated with the presence of such strong-zero mode operator when in non-equilibrium.

This strong zero mode operator is immune to decoherence~\cite{Chamon}, potentially leading to topological qubits with infinite coherence time~\cite{Laflorencie} 
This occurs because they are topologically protected due to global parity conservation and  quasi-degenerate paired states in the full spectrum
(involving opposite fermion parity)~\cite{Akhmerov,Kells}. 
The spectral weight is dominated by only a few  higher energy degenerate-pair states, all with comparable weight,
 contributing to $\omega=e_n-e_m=0$ (with $|m-n|=1$) in the $LDOS_{elec}(\omega,t)$ (Eq.~6) and $LDOS_{hole}(\omega,t)$. 
%
This protection
survives as long as the left and right MZMs do not overlap with each other. 
However, increasing $V$, the peaks of $LDOS_{elec}(\omega,t)$ and $LDOS_{hole}(\omega,t)$ start splitting and
the peak values are no longer equal in magnitude (Fig.~\ref{Fig4}(c)).
 Furthermore, at large $V$ the electron and hole parts of LDOS
show peaks away from zero and these peaks are largely splitted. The split in  $LDOS_{elec}(\omega,t)$ and $LDOS_{hole}(\omega,t)$
peaks is due to the breakdown of degeneracy of higher excited paired states  
increasing $V$.

{ \it Fast Fusion of Majoranas.}
In real Majorana nanowire setups,
it is necessary to move the Majoranas with sufficient speed to be faster than the quasi-particle ``poisoning'' 
time)~\cite{Aasen,Marcus,Rainis,Chamon1}. 
Here, we present the minimum required switching time of local gates for fast moving MZM$^R$, so that
we obtain a full electron after the fusion of left and right MZMs. As discussed earlier, to fuse Majoranas 
we moved at various speeds the right MZM$^R$ all the way to the left end (site $j=1$). 
In Figs.~\ref{Fig5}(a,b,c,d), we show the fusion of MZMs using a potential Wall ($\mu(t)>0$) with initial state
having parity $P=-1$ and for $\Delta=1.0$ and at $V=0$ (Figs.~\ref{Fig5}(a,b)) and $V=2$ (Figs.~\ref{Fig5}(c,d)).

For $V=0$ ($\Delta = 1.0$), the many-body eigenstates of $H(t)$ have degenerate paired states.
For this reason, even for smaller  $\tau$ (faster-motion of MZM$^R$), we have well-defined sharp peaks at $\omega=0$ in 
$LDOS_{elec}(\omega)$ and $LDOS_{hole}(\omega)$ at $V=0$ (see Figs.~\ref{Fig5}(a,b)).
Moreover, for $LDOS_{hole}(\omega,t)$ the spectral weight only arises from 
higher energy degenerate-pair states [$m,n \ge 3$]. 
Increasing $\tau$ (slower-motion of MZM$^R$),
the peak at $\omega=0$ for  $LDOS_{elec}(\omega)$ starts increasing,
while the $LDOS_{hole}(\omega)$ peak value at $\omega=0$ decreases.
At $\tau=36$ (slow), we obtain a sharp electron peak for $V=0$ close to $\omega=0$ (Figs.~\ref{Fig5}(a)),
whereas, there is no peak for the hole $LDOS_{hole}(\omega)$ (Figs.~\ref{Fig5}(b)) at $\omega=0$, 
confirming the formation of a full electron (for $\tau \ge 36$).
 For $\tau \ge 36$,  the contribution in the $LDOS_{elec}(\omega)$ peak at $\omega=0$ arises only from $m,n=1$ or $2$ in Eq.~6.

However, at $V=2$ the degeneracy of bulk excited states no longer 
exists (although states close to ground-states are still nearly degenerate).
This leads to asymmetrically smeared peaks 
for $LDOS_{elec}(\omega)$ and $LDOS_{hole}(\omega)$ around $\omega=0$ at
small values of $\tau$ and strong Coulomb interaction $V$ (Figs.~\ref{Fig5}(c,d)).
For $V=2$, it is required a slower rate of movement $\tau=60$, as compared to $V=0$,
to form a full electron after Majorana fusion (Fig.~\ref{Fig5}(c)).
For example, at $\tau \ge 60$ the contribution to the $LDOS_{elec}(\omega)$ peak at $\omega=0$ arises only from $m,n=1$ or $2$ in Eq.~6.
Then, clearly the Coulomb repulsion {\it reduces} the topological protection by reducing the effective gap between the lower
state manifold and the rest of the states. Thus, as $V$ grows a much slower movement of MZM$^R$ is needed to reach adiabaticity.

In Figs.~\ref{Fig5}(e,f), we show the fusion of Majoranas using a potential Well ($\mu(t)<0$) now for $\Delta=0.6$, $V=1$, and increasing
the length to $L=16$, 
with initial state having parity $P=+1$. For these parameters, we believe our system is closer to realistic setups~\cite{Herbrych}
for experiments~\cite{Yazdani,jeon}. However, for negative chemical potential, after the formation of an
 electron at site $j=1$ the repulsive nearest-neighbor $V$ leads to the split in ground state energy (approximately of order $V$~\cite{comment}),
causing an energy shift in peak values of $LDOS_{elec}(\omega)$ and $LDOS_{hole}(\omega)$ increasing $\tau$. For example,
at $\tau=48$, we obtain a sharp electron peak close to $\omega=1.0$, whereas  the $LDOS_{hole}(\omega)$ peak vanishes to zero as $\tau$ grows (Figs.~\ref{Fig5}(e,f)).
Specifically, this shows the formation of a full electron at $\tau=48$ for  $\Delta=0.6$, $V=1$, and $L=16$.

In terms of SI units the switching time  for $V=0$ corresponds to $\tau \hbar/\Delta\sim 0.13$ns to $3.9$ns (using $\tau=36$, and
$\Delta=180 \mu$eV  or $\Delta=6 \mu$eV as in previous literature~\cite{Zhou,Marcus}). 
This is the time required per gate, which, as example, was five in~\cite{Zhou}. 
Independently, the quasiparticle ``poisoning'' time in nano-wire systems has been 
estimated in a broad range $10$ns to $10$ms~\cite{Rainis,Marcus}. Because in the worse case of $3.9$ns, five gates require a total time $19.5$ns
to move adiabatically the Majorana, and since this number is very close to the lower bound of poisoning time, we conclude that 
there should be a time range where moving Majoranas in chains can occur adiabatically before poisoning occurs for $V=0$. As $V$ increases,
the situation deteriorates because at $V=2$ we must use $\tau=60$,
 but the new adiabatic time needed is only $33$ns, still close to the experimental lower poisoning time.

{ \it Conclusions. } 
We performed real-time dynamics and fusion of Majoranas in the interacting 1D Kitaev model 
using sequential application of time-dependent chemical potentials (gates).
We show that the movement and fusion outcomes can be monitored using the time-dependent local density-of-states,
and should be observed in tunneling spectroscopy experiments~\cite{Zhang}.
We find that for non-interacting and fast moving Majoranas, the near degeneracy of MZMs exists even in many higher energy states and MZMs remain
topologically protected. However,  for the interacting case, with increasing $V$ we find a decrease in
spectral weight at $\omega=0$ in the time-dependent local density-of-states
 (with non-equal and splitted peaks of electron and hole component of $LDOS(\omega,j,t)$). 
Furthermore, we estimate the minimum required switching time of local gates 
to form a full electron  after the fusion. Due to advancements in fabricating Majorana nanowires~\cite{Madsen,Zhang} with long quasiparticle poisoning time~\cite{Marcus,Rainis}, and considering our estimations for 
the times needed for adiabatic movement,
we believe proper Majorana movement could be realized in realistic gate-control nanowire devices~\cite{Zhang,Chetan2,comment2}.


{\it Acknowledgments.} 
B.P., N.M. and E.D. were supported by the U.S. Department of
Energy (DOE), Office of Science, Basic Energy Sciences
(BES), Materials Sciences and Engineering Division.


\begin{thebibliography}{99}
\bibitem{foot} The value 0.48 instead of 0.5 of spectral weight might arise from an integration error due to
                   finite $\omega$ steps and the broadening $\eta$ used.
\bibitem{foot2}{For $t_h=\Delta=1$, the system has many states with the same energy above the gap and
             we are not able to calculate the parity of these high-energy states. However, the electron and hole components of $LDOS(\omega,j,t)$ 
             remain the same for fast moving Majoranas.}
\end{thebibliography}
\end{document}